# Finite Element Modeling of Charge and Spin-currents in Magnetoresistive Pillars with Current Crowding Effects

N. Strelkov[1,2], A. Vedyayev[1,2], D. Gusakova[1,3,a)], L. D. Buda-Prejbeanu[1], M. Chshiev[1], S. Amara[1], A. Vaysset[1], B. Dieny[1]

[1]*SPINTEC, UMR-8191, CEA-INAC/ CNRS/UJF-Grenoble 1/Grenoble-INP, 17 rue des martyrs, 38054 GRENOBLE cedex 9*
[2]*Lomonosov University, Faculty of Physics, Department of Magnetism, Moscow, Russia*
[3]*CEA-LETI, MINATEC, DRT/LETI/DIHS, 38054 Grenoble, France*

**The charge and spin diffusion equations taking into account spin-flip and spin-transfer torque were numerically solved using a finite element method in complex non-collinear geometry. This approach was used to study the spin-dependent transport in giant magnetoresistance metallic pillars sandwiched between extended electrodes as in magnetoresistive heads for hard disk drives. The charge current crowding around the boundaries between the electrodes and the pillar has a quite significant influence on the spin current.**

Spin electronics was born in 1988 with the discovery of Giant Magnetoresistance (GMR)[1,2]. Since then, it has been expanding thanks to a strong synergy between fundamental research and industrial developments particularly concerning magnetoresistive heads for hard disk drives[3,4], Magnetic Random Access Memories (MRAM)[5], logic devices[6] and RF oscillators[7]. Most of these spintronic devices under research and development involve inhomogeneous current flows. This is the case in metallic pillars or low resistance tunnel junctions implying current crowding effects[8], in point contact RF oscillators[9], in GMR current-perpendicular-to-plane (CPP) magnetoresistive heads and especially in current confined path (CCP) structures[10,11]. To quantitatively interpret experimental data in these complex geometries or to design spintronic devices with non-uniform current flow, it is therefore important to develop a theoretical tool which is able to describe the spin-dependent transport (charge and spin-currents) as well as spin-transfer torque in systems of arbitrary shape and magnetic configuration.

The purpose of the present study was to develop such a tool in the case of diffusive transport. In this letter, we present our approach to accomplish this goal and apply it to the calculation of the transport properties in CPP-GMR pillars sandwiched between extended

[a)] Corresponding author: daria.gusakova@cea.fr

electrodes. We show that the current crowding effect which takes place in such structure gives rise to quite peculiar spin transport phenomena.

The general formalism that we used in the diffusion limit is derived based on the extension of the Valet and Fert theory[12,13]. Each material constituting the system of arbitrary shape and composition is described by local transport parameters ($C_0$–conductivity, $\beta$–spin asymmetry of $C_0$, $D_0$–diffusion constant, $\beta'$–spin asymmetry of $D_0$, $N_0$–density of states at Fermi level).

In this study we assume $\beta=\beta'$ in the following. All transport properties are then described by four local variables: the scalar electrostatic potential $\tilde{\varphi}$ and the 3 components of spin accumulation in spin-space $(m_x, m_y, m_z)$. The local charge current vector is then given by:

$$\mathbf{j}^e = \frac{2C_0}{e}\nabla\tilde{\varphi} - \frac{2\beta C_0}{e^2 N_0}(\mathbf{u}_M, \nabla\mathbf{m}) \qquad (1),$$

where $\mathbf{u}_M$ is a unit vector parallel to the local magnetization. e is the electron charge. The spin current is described by a tensor (3 coordinates in spin space, 3 coordinates in real space) and expressed as:

$$\mathbf{j}^m = \frac{2\beta C_0}{e}\mathbf{u}_M \nabla\tilde{\varphi} - \frac{2C_0}{e^2 N_0}\nabla\mathbf{m} \qquad (2).$$

The four variables are then calculated everywhere in space in steady state by numerically solving the two fundamental equations of spin-dependent diffuse transport (4 scalar equations):

$$\begin{cases} div\mathbf{j}^e = 0 & (3) \\ div\mathbf{j}^m + \frac{J_{sd}VM_S}{\hbar\mu_B}(\mathbf{m}\times\mathbf{u}_M) + \frac{\mathbf{m}}{\tau_{sf}} = 0 & (4) \end{cases}$$

where $J_{sd}$, $M_S$, $V$ and $\tau_{sf}$ represent the *s-d* exchange interaction constant, the saturation magnetization, the volume and the spin relaxation time, respectively; $\hbar$ and $\mu_B$ are the Planck constant and the Bohr magneton.

The first equation expresses the conservation of charge. The second one states that the spin polarization of the current is not conserved. It can vary either because of spin relaxation or because of the local spin-transfer torque which induces a precession of the spin accumulation around the local magnetization due to *s-d* exchange interaction. The spin-transfer torque is itself given by $\mathbf{T} = \frac{J_{sd}VM_S}{\mu_B}(\mathbf{m}\times\mathbf{u}_M)$.

The constants $J_{sd}$ and $\tau_{sf}$ are related to two characteristic lengths by $\lambda_J = \sqrt{2\hbar D_0 / J_{sd}}$ where $\lambda_J$ is the spin-reorientation length, i.e. the distance over which the spin polarization gets reoriented along the local magnetization and $l_{sf} = \sqrt{2(1-\beta^2)D_0\tau_{sf}}$ where $\tau_{sf}$ is the spin-flip relaxation time. In addition, at outer boundaries, we impose no perpendicular component of charge and spin current except at the boundaries where a potential is applied.

Using this general formalism adapted for the finite element solver, the spatial distribution of the charge and spin currents as well as the spin transfer torque was calculated in two dimensional magnetoresistive nanopillars sandwiched between two non-magnetic extended metallic electrodes as shown in Fig.1(a). The nanopillar consists of two 3nm thick magnetic layers (reference and free layers) separated by a 2nm thick non-magnetic metallic spacer. We assume that the relative orientation of the magnetizations in the two magnetic layers can be varied in the plane perpendicular to *x*-axis. Voltages of respectively $\varphi_{in} = 0V$ ($\varphi_{out} = 50mV$) are uniformly applied between the top surfaces of the two electrodes. In this model study, we only considered bulk spin-dependent scattering and bulk spin relaxation. The bulk parameters that we used for the various materials are representative for the case of Co and Cu[14]. Under these assumptions, the resistance of the stack in parallel (antiparallel) configuration is found to be $R_P$=401Ω ($R_{AP}$=403Ω), yielding a magnetoresistance $\Delta R/R_P$=0.5%.

Fig.1(b) shows the charge current flow through the device (arrows) and the charge current amplitude (color map) throughout the structure in parallel magnetic configuration. Because the charge current is trying to follow the shortest path throughout the structure, it has a pronounced vertical gradient within the magnetoresistive pillar. The current amplitude is actually ten times larger in the upper part of the pillar than in its bottom part. Hot spots are visible around the upper corners of the pillar.

Fig. 2 shows the spin-current distribution (arrows) of the spin component parallel to the *y*-axis, i.e. to the magnetization of the reference layer, for two magnetic configurations: parallel (Fig.2(a)) and antiparallel (Fig.2(b)). The color map represents the *y*-component of the spin accumulation.

In the parallel case (Fig.2(a)) the electrons are initially unpolarized in the Cu electrodes and get more and more polarized as they approach the magnetic pillar. Since the charge current is much more intense in the upper part of the pillar compared to that in the lower part, the y-component of spin current is also much more intense in the upper part of the pillar than in the lower part. Correlatively, a large excess of electron spin aligned with negative *y*-axis arises in the Cu electrode on the left-hand side of the pillar. Symmetrically, an excess of electron spins aligned with positive *y*-axis appears in the Cu electrode on the right-hand side of the pillar. These excess densities of polarized electrons are largest on both sides of the upper part of the pillar as indicated in Fig.2(a) and are minimal around the bottom part of the structure. As a result of this distribution of spin accumulation, vortex of diffusion spin current is formed as represented by the white arrows in Fig.2(a). Unexpectedly, this implies that the *y*-component of spin current has opposite directions in the upper and lower parts of the magnetoresistive pillar.

The situation in antiparallel configuration is quite different (Fig.2(b)). The *y*-component of spin accumulation now has maximum in the non-magnetic spacer layer but is much larger in the upper part of the pillar than in its lower part because of the vertical current density gradient. The resulting gradient of *y*-component of spin accumulation gives rise to an intense in-plane *y*-component of spin current flowing in the spacer layer as well as two symmetric vortices of *y*-component of spin current flowing in the Cu electrodes on both sides of the pillar (as indicated by white arrows in Fig.2(b)).

Fig. 3 shows the reduced resistance $r = [R(\theta) - R(0)]/[R(\pi) - R(0)]$ versus the angle between the magnetizations of the two magnetic layers in two situations: i) the present nanopillar sandwiched between two extended electrodes and ii) the same nanopillar sandwiched between two electrodes extending along the x direction and having the same diameter as the pillar so that the charge current is uniform throughout the stack (1-dimensional model with charge current flowing along *x*-axis).

Clearly, the angular variation of CPP-GMR does not follow a simple cosine variation which is in agreement with theoretical expectation[15]. Furthermore, it follows from Fig. 3 that the system geometry influences the angular variation. This further emphasizes the need to take into account the influence of spatial current non uniformities in the design of spintronic devices.

As a further step, we show in Fig. 4(a) and Fig. 4(b) the in-plane (Slonczewski's term[16]) and perpendicular-to-plane (field-like term[13]) components of spin transfer torque (STT) in 90° magnetic configuration. The black arrows represent the *y*-component of spin current whereas the color map is related to the STT amplitude. As already pointed out for metallic pillars[13], the perpendicular component of the torque is much smaller (by more than 1 order of magnitude) than the in-plane component. Furthermore, since the charge current density is much higher in the upper part of the pillar than in its bottom part, the STT amplitude is also much larger in the upper part of the pillar. From the viewpoint of magnetization dynamics, this implies that magnetic excitations due to STT are likely to be first generated in the upper part of the pillar as the current density is increased above the STT excitation threshold.

In conclusion, a numerical tool has been developed to compute the charge and spin-current in magnetic structures of arbitrary shape and composition. This tool has been used to investigate the spin-dependent transport properties through magnetoresistive nanopillars sandwiched between extended electrodes. It was shown that the current crowding effect gives rise to strong in-plane inhomogeneities in spin accumulation yielding large in-plane components of spin-current. This type of tool should be quite helpful in the design of functional spintronic devices as well as for the quantitative interpretation of experimental data in devices with non uniform or non-local currents. As a next step, the computation of the micromagnetic dynamics will be self-consistently coupled to these calculations of transport properties.

This project has been supported in parts by the European RTN "Spinswitch" MRTN-CT-2006-035327 and partially by Chair of Excellence Program of the Nanosciences Foundation (Grenoble, France). D.G. acknowledges the French National Research Agency (ANR) (CARNOT program).

FIGURES CAPTIONS

FIG. 1. (color online) (a) Scheme of studied magnetoresistive nanopillar sandwiched between two extended electrodes. The pillar composition is representative of Co3nm/Cu2nm/Co3nm. (b) Zoom on the magnetoresistive nanopillar showing the charge current flow (arrows) and charge current amplitude (color map).

FIG. 2. (color online) Zoom around the magnetoresistive pillar showing the $y$-component of spin current flow throughout the system in (a) parallel magnetic configuration, (b) antiparallel configuration. The normalized arrows indicate the spin current flow whereas the color map represents the $y$-component of the spin accumulation.

FIG. 3. Comparison of angular variation of CPP-GMR in the present nanopillar sandwiched between extended electrode (black squares) and throughout the same nanopillar assuming uniform current flow in $x$ direction -1D model (grey circles).

FIG. 4. (a) In-plane and (b) perpendicular-to-plane components of spin transfer torque in 90° magnetic configuration. The black arrows represent the $y$-component of spin current whereas the color map is associated with the STT amplitude.

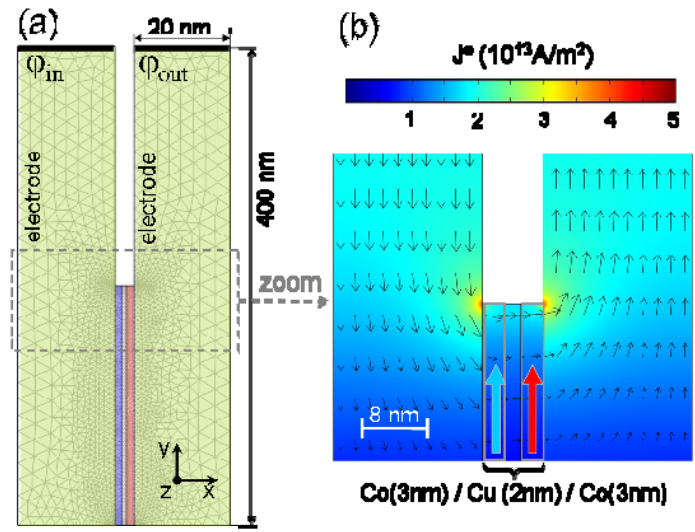

FIG. 1.　　　　　　　　　　　　　　N. Strelkov *et al*

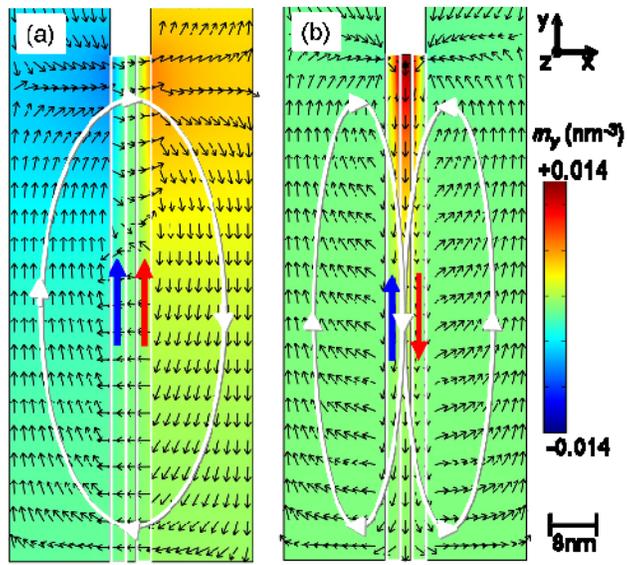

FIG. 2. N. Strelkov *et al*

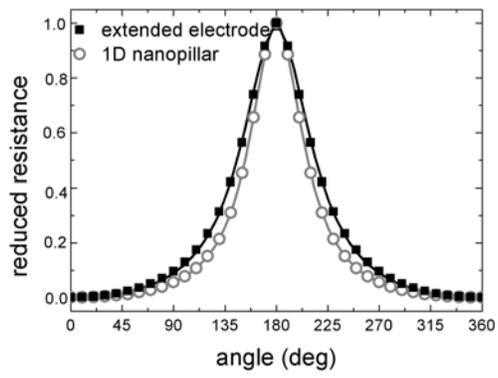

FIG. 3. N. Strelkov *et al*

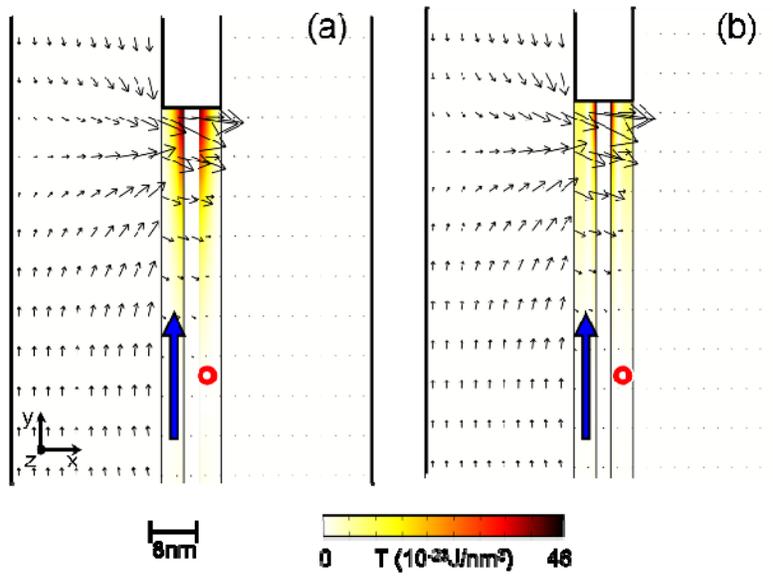

FIG. 4.                                                        N. Strelkov *et al*